\documentclass[amsmath,11pt]{article}

\date{\empty}

\begin{document}

\title{\bf Gravito-electromagnetic resonances}

\author{Christos G. Tsagas\\ {\small Section of Astrophysics,
Astronomy and Mechanics, Department of Physics}\\ {\small Aristotle
University of Thessaloniki, Thessaloniki 54124, Greece}}

\maketitle

\begin{abstract}
The interaction between gravitational and electromagnetic radiation has a rather long research history. It is well known, in particular, that gravity-wave distortions can drive propagating electromagnetic signals. Since forced oscillations provide the natural stage for resonances to occur, gravito-electromagnetic resonances have been investigated as a means of more efficient gravity-wave detection methods. In this report, we consider the coupling between the Weyl and the Maxwell fields on a Minkowski background, which also applies to astrophysical environments where gravity is weak, at the second perturbative level. We use covariant methods that describe gravitational waves via the transverse component of the shear, instead of pure-tensor metric perturbations. The aim is to calculate the properties of the electromagnetic signal, which emerges from the interaction of its linear counterpart with an incoming gravitational wave. Our analysis shows how the wavelength and the amplitude of the gravitationally driven electromagnetic wave vary with the initial conditions. More specifically, for certain initial data, the amplitude of the induced electromagnetic signal is found to diverge. Analogous, diverging, gravito-electromagnetic resonances were also reported in cosmology. Given that, we extend our Minkowski-space study to cosmology and discuss analogies and differences in the physics and in the phenomenology of the Weyl-Maxwell coupling between the aforementioned two physical environments.
\end{abstract}

\section{Introduction}\label{sI}
Gravitational and electromagnetic waves are believed to interact in a variety of ways. The literature contains a number of studies showing how gravitational radiation affects the propagation of electromagnetic signals by modifying their amplitude, their wavelength, their polarisation and their direction, either in vacuum or in the presence of conductive plasmas (e.g.~see~\cite{Z}). There are also solutions of Einstein's equations that describe colliding plane gravitational and electromagnetic waves on a flat Minkowski background. Only recently, however, techniques that could construct interacting solutions from the wave parameters before the collision were introduced (see~\cite{AG} and references therein). Historically, most of the interest in the coupling between the Weyl and the Maxwell fields has been motivated by the on going search for gravitational-wave signals and by the quest for new and potentially more efficient detection mechanisms~\cite{BM}. Several of the aforementioned references raise the possibility of resonances between the two sources and then exploit it, either as a novel gravity-wave detection method~\cite{BM}, or as a general relativistic mechanism of amplifying large-scale magnetic fields~\cite{T1}.

Gravitational waves can affect electromagnetic signals directly (e.g.~by modifying their amplitude, wavelength, etc) as well as indirectly (e.g.~by altering the geometry, or the nature, of the host environment)~\cite{T}. Here, we are interested in the direct effects, with particular emphasis on the possibility of resonances. The gravito-electromagnetic interaction is studied at second order on an empty, static Minkowski background. This means that our results apply to environments where the gravitational field is weak, like the interstellar space or the laboratory for example. In contrast to the majority of the approaches, which are metric based, we employ the 1+3 covariant approach to General Relativity~\cite{TCM}. The main difference is in the description of the gravitational radiation, which covariantly is monitored through the electric and magnetic components of the Weyl tensor~\cite{H}. The latter determines the free (i.e.~the long range) gravitational field, the transverse part of which corresponds to gravity waves.

Weyl curvature distortions affect the propagation of electromagnetic signals via Maxwell's equations. Gravitational waves trigger (pure-tensor) shear perturbations that directly affect both the electric and the magnetic component of the signal, through (the generalised) Ampere's and Faraday's laws. On a Minkowski background, the two sources propagate as simple (linear) plane waves, while their interaction is monitored by a second-order system that describes the gravitationally induced electromagnetic signal. Not surprisingly, the latter differs from the original one and the differences depend on the specifics of the gravito-electromagnetic interaction. For our purposes, however, the most important feature of the aforementioned second-order system is that it describes two forced oscillations (one for each component of the electromagnetic signal). The importance lies in the long known fact that forced vibrations provide the natural physical environment for resonances to occur. This means that, in principle at least, the interaction between gravitational and electromagnetic waves can lead to the resonant amplification of the latter.

Although resonances can occur spontaneously, they require rather special conditions. More specifically, the driving and the driven wave must oscillate in tune. In our case, the driving force of the gravitationally induced electromagnetic signal is described as a superposition of two waves. The characteristics of the latter,  their effective wavelengths in particular, are fixed at the beginning of the interaction. The same initial conditions also determine the wavelength of the gravitationally induced -- the driven -- electromagnetic signal. We provide a set of formulae that allow one to calculate the specifics of the induced electromagnetic wave for a variety of initial conditions. We find that the wavelengths of the driving and the driven waves are generally different, in which case the resulting electromagnetic oscillation will experience mild amplification (if any). When the circumstances are favourable, however, these two wavelengths will coincide and the driving and the driven source will oscillate in resonance. It is then, that the amplitude of the gravitationally induced electromagnetic signal is found to diverge. Whether this happens or not, according to our analysis, depends also on the interaction angle between the original electromagnetic and gravitational waves. Here, we find that Weyl-Maxwell resonances occur when the aforementioned two interacting waves propagate in the same, or in the opposite, direction.

These results should not come as a complete surprise. After all, we were dealing with forced oscillations, where resonances naturally occur. In fact, gravito-electromagnetic resonances have a rather long history in the relevant literature and certain types of them have been proposed in the past as efficient mechanisms of gravitational-wave detection. To the best of our knowledge, however, the first time such strong (diverging) resonances were reported in the literature was in the context of cosmology~\cite{T1}. Given that, and also for reasons of completeness and comparison, we extend our analysis to include cosmological environments as well. In cosmology, the host spacetime is no longer static, empty or flat. This means that, although the basic physics of the gravito-electromagnetic interaction remain essentially the same, the specifics change. The main difference appears to come from the expansion, which adds a ``friction'' term to the wave-equations that generally makes the resonance peak narrower.

\section{Electromagnetic and gravitational waves in Minkowski 
space}\label{sEMGWMS}
In weak-gravity environments, for example in the interstellar space (away from massive compact stars) or in the laboratory, the spacetime is effectively flat. There, one can use the Minkowski metric to describe the geometry of the host environment.

\subsection{Electromagnetic waves}\label{ssEMWs}
Consider electromagnetic radiation traveling on an empty, static Minkowski background. The Maxwell field is invariantly described by means of the antisymmetric Faraday tensor ($F_{ab}$), which splits into its electric and magnetic components once a family of timelike observers is introduced. Assuming that $u_a$ is the 4-velocity of the aforementioned observers (normalised so that $u_au^a=-1$), the corresponding electric and magnetic fields are $E_a=F_{ab}u^b$ and $B_a=\varepsilon_{abc}F^{bc}/2$ respectively ($\varepsilon_{abc}$ is the Levi-Civita tensor of the observers' 3-dimensional space). On our Minkowski background, the two fields obey the linear wave equations~\cite{T2}
\begin{equation}
\ddot{E}_a- {\rm D}^2E_a= 0 \hspace{15mm} {\rm and} \hspace{15mm} \ddot{B}_a- {\rm D}^2B_a= 0\,,  \label{lEM}
\end{equation}
with the overdots indicating proper-time derivatives along the observers' worldlines. Also, ${\rm D}^2={\rm D}^a{\rm D}_a$ represents the covariant Laplacian operating on the observers' 3-dimensional rest-space.\footnote{Given the flatness of the background metric, we have ${\rm D}_a\rightarrow\partial_a$ and ${\rm D}^2={\rm D}^a{\rm D}_a\rightarrow\partial^a\partial_a= \partial^2$ to zero order.}

Equations (\ref{lEM}a) and (\ref{lEM}b) are straightforward to solve. To proceed, we decompose the components of the electromagnetic signal by introducing the familiar harmonic splitting
\begin{equation}
E_a= \sum_{n}E_{(n)}\mathcal{Q}_a^{(1)(n)} \hspace{15mm} {\rm and} \hspace{15mm} B_a= \sum_{n}B_{(n)}\mathcal{Q}_a^{(2)(n)}\,,  \label{EMsplit}
\end{equation}
where $\mathcal{Q}_a^{(1)(n)}$, $\mathcal{Q}_a^{(2)(n)}$ are the corresponding vector harmonics and $n$ is the eigenvalue/wavenumber of the associated electromagnetic mode. Note that $n^2=n^an_a$, with $n_a$ representing the wavevector along the propagation direction of the signal. Also, by construction, ${\rm D}_aE_{(n)}=0={\rm D}_aB_{(n)}$, $\dot{\mathcal{Q}}_a^{(1)(n)}=0= \dot{\mathcal{Q}}_a^{(2)(n)}$, ${\rm D}^a\mathcal{Q}_a^{(1)(n)}= 0={\rm D}^a\mathcal{Q}_a^{(2)(n)}$ and ${\rm D}_{\langle b}\mathcal{Q}^{(1)(n)}_{a\rangle}=0={\rm D}_{\langle b}\mathcal{Q}^{(2)(n)}_{a\rangle}$.\footnote{Angled brackets indicate the symmetric and trace-free component of second-rank spacelike tensors.} Finally, we point out that both of the harmonic functions satisfy the vector version of the Laplace-Beltrami equation, that is
\begin{equation}
{\rm D}^2\mathcal{Q}_a^{(1)(n)}= -n^2\mathcal{Q}_a^{(1)(n)} \hspace{15 mm} {\rm and} \hspace{15mm} {\rm D}^2\mathcal{Q}_a^{(2)(n)}= -n^2\mathcal{Q}_a^{(2)(n)}\,. \label{vLB}
\end{equation}
Substituting decompositions (\ref{EMsplit}a) and (\ref{EMsplit}b) back into Eqs.~(\ref{lEM}a) and (\ref{lEM}b) and then using expressions (\ref{vLB}), the original wave formulae recast into
\begin{equation}
\ddot{E}_{(n)}+ n^2E_{(n)}= 0 \hspace{15mm} {\rm and} \hspace{15mm} \ddot{B}_{(n)}+ n^2B_{(n)}= 0\,,  \label{dlEM}
\end{equation}
which now monitor the linear evolution of the signals' $n$-th harmonic mode. Both of the above are straightforward to solve, leading to the solution
\begin{eqnarray}
E_{(n)}&=& \left[E_0^{(n)}\sin\left(nt_0\right) +{1\over n}\, \dot{E}_0^{(n)}\cos\left(nt_0\right)\right]\sin\left(nt\right) \nonumber\\&&+\left[E_0^{(n)}\cos\left(nt_0\right) -{1\over n}\, \dot{E}_0^{(n)}\sin\left(nt_0\right)\right]\cos\left(nt\right)\,,  \label{lEn}
\end{eqnarray}
for the electric component of the Maxwell field, and to solution
\begin{eqnarray}
B_{(n)}&=& \left[B_0^{(n)}\sin\left(nt_0\right) +{1\over n}\, \dot{B}_0^{(n)}\cos\left(nt_0\right)\right]\sin\left(nt\right) \nonumber\\&&+\left[B_0^{(n)}\cos\left(nt_0\right) -{1\over n}\, \dot{B}_0^{(n)}\sin\left(nt_0\right)\right]\cos\left(nt\right)\,,  \label{lBn}
\end{eqnarray}
for its magnetic counterpart. This set monitors the propagation of an electromagnetic wave on an empty, static Minkowski background to first order. Note that the zero suffix indicates a given initial time, at which both the amplitude and the phase of signal are determined.

\subsection{Gravitational waves}\label{ssGWs}
On the same Minkowski space, we will also consider the propagation of gravitational radiation. Covariantly, gravitational waves are described by means of the transverse parts of the electric and magnetic components of the Weyl field (e.g.~see~\cite{H}). Nevertheless, given the symmetries of our background and the absence of matter, linear Weyl curvature distortions are monitored by the pure-tensor modes of the shear.\footnote{Also true for gravity waves propagating on a Friedmann-Robertson-Walker (FRW) background (see \S~\ref{ssBLDs}).} These are isolated by imposing the first-order constraint ${\rm D}^b\sigma_{ab}=0$ (e.g.~see~\cite{TCM}), which in our case is trivially satisfied at all times. Moreover, on a Minkowski background, the transverse component of the shear obeys the linear wave equation
\begin{equation}
\ddot{\sigma}_{ab}- {\rm D}^2\sigma_{ab}= 0\,.  \label{lGW}
\end{equation}
As before, the overdots indicate time-differentiation relative to the fundamental observers and ${\rm D}^2={\rm D}^a{\rm D}_a$ represents their associated 3-D Laplacian operator. We also introduce the harmonic splitting
\begin{equation}
\sigma_{ab}= \sum_k\sigma_{(k)}\mathcal{Q}_{ab}^{(k)}\,,  \label{GWsplit}
\end{equation}
where $\sigma_{(k)}$ is the $k$-mode of the shear perturbation and ${\rm D}_a\sigma_{(k)}=0$. Note that $k^2=k^ak_a$, with $k$ being the mode's eigenvalue/wavenumber and $k_a$ the corresponding wavevector. Also, $\mathcal{Q}_{ab}^{(k)}$ are tensor harmonic functions, which satisfy the conditions $\mathcal{Q}_{ab}^{(k)}=\mathcal{Q}_{\langle ab\rangle}^{(k)}$, $\dot{\mathcal{Q}}_{ab}^{(k)}=0={\rm D}^b\mathcal{Q}_{ab}^{(k)}$ and the tensor form
\begin{equation}
{\rm D}^2\mathcal{Q}_{ab}^{(k)}= -k^2\mathcal{Q}_{ab}^{(k)}\,,  \label{tLB}
\end{equation}
of the Laplace-Beltrami equation. Combining expressions (\ref{lGW}), (\ref{GWsplit}) and (\ref{tLB}) leads to the linear wave equation
\begin{equation}
\ddot{\sigma}_{(k)}+ k^2\sigma_{(k)}= 0\,,  \label{dlGW}
\end{equation}
which monitors the $k$-th harmonic mode of a gravitational wave propagating on an empty, static Minkowski background. As it is straightforward to show, the above accepts the solution
\begin{eqnarray}
\sigma_{(k)}&=& \left[\sigma_0^{(k)}\sin\left(kt_0\right)+{1\over k}\, \dot{\sigma}_0^{(k)}\cos\left(kt_0\right)\right] \sin\left(kt\right) \nonumber\\ &&+\left[\sigma_0^{(k)} \cos\left(kt_0\right)-{1\over k}\,\dot{\sigma}_0^{(k)} \sin\left(kt_0\right)\right]\cos\left(kt\right)\,,  \label{lGWk}
\end{eqnarray}
with the zero suffix corresponding to a given initial time and the quantities in brackets determining the characteristics of the wave (i.e.~its amplitude and phase).

\section{Gravito-electromagnetic interaction in Minkowski
space}\label{sGEMIMS}
Technically speaking, gravitational waves affect the propagation of their electromagnetic counterparts through Maxwell's equations. These combine to provide a set of two wave-equations (one for each component of the electromagnetic field), which allow us to study the coupling between the two sources at different perturbative levels.

\subsection{Second-order equations}\label{ssSoEs}
The last two sections considered the linear evolution of electromagnetic and gravitational radiation on an empty, static Minkowski background. Here, we will look into the nonlinear (second-order in particular) interaction of these two sources. Our aim is to study the effect of gravitational waves on electromagnetic signals. At the aforementioned perturbative level, the interaction between the Weyl and the Maxwell fields is governed by the system~\cite{T2}
\begin{equation}
\ddot{E}_a- {\rm D}^2E_a= \sigma_{ab}\dot{\tilde{E}}^b+ \varepsilon_{abc}\tilde{B}_d{\rm D}^b\sigma^{cd}- 2\varepsilon_{abc}\sigma^b{}_d{\rm D}^{\langle c}\tilde{B}^{d\rangle}- \mathcal{R}_{ab}\tilde{E}^b- E_{ab}\tilde{E}^b+ H_{ab}\tilde{B}^b  \label{soG-E1}
\end{equation}
and
\begin{equation}
\ddot{B}_a- {\rm D}^2B_a= \sigma_{ab}\dot{\tilde{B}}^b- \varepsilon_{abc}\tilde{E}_d{\rm D}^b\sigma^{cd}+ 2\varepsilon_{abc}\sigma^b{}_d{\rm D}^{\langle c}\tilde{E}^{d\rangle}- \mathcal{R}_{ab}\tilde{B}^b- E_{ab}\tilde{B}^b- H_{ab}\tilde{E}^b\,,  \label{soG-B1}
\end{equation}
where $\tilde{E}_a$, $\tilde{B}_a$ and $E_a$, $B_a$ are the electromagnetic components before and after the interaction respectively. The former obey the simple, linear wave solutions (\ref{lEn}) and (\ref{lBn}) given in \S~\ref{ssEMWs}. Similarly, the transverse part of the shear satisfies solution (\ref{lGWk}). We also note that $R_{ab}$ is the perturbed (linear) 3-Ricci tensor, while $E_{ab}$ and $H_{ab}$ represent the (linearised on our Minkowski background) electric and magnetic parts of the Weyl tensor respectively. Then, to first order, we have~\cite{TCM}
\begin{equation}
\mathcal{R}_{ab}= E_{ab}\,, \hspace{15mm} E_{ab}= -\dot{\sigma}_{ab} \hspace{15mm} {\rm and} \hspace{15mm} H_{ab}= {\rm curl}\sigma_{ab}  \label{lEqs}
\end{equation}
with ${\rm curl}\sigma_{ab}=\varepsilon_{cd\langle a}{\rm D}^{c}\sigma^d{}_{b\rangle}$ by definition. Substituting the auxiliary linear relations (\ref{lEqs}) into Eqs.~(\ref{soG-E1}) and (\ref{soG-B1}), the latter recast into
\begin{equation}
\ddot{E}_a- {\rm D}^2E_a= \sigma_{ab}\dot{\tilde{E}}^b+ 2\dot{\sigma}_{ab}\tilde{E}^b+ \varepsilon_{abc}\tilde{B}_d{\rm D}^b\sigma^{cd}- 2\varepsilon_{abc}\sigma^b{}_d{\rm D}^{\langle c}\tilde{B}^{d\rangle}+ \tilde{B}^b{\rm curl}\sigma_{ab}  \label{soG-E2}
\end{equation}
and
\begin{equation}
\ddot{B}_a- {\rm D}^2B_a= \sigma_{ab}\dot{\tilde{B}}^b+ 2\dot{\sigma}_{ab}\tilde{B}^b- \varepsilon_{abc}\tilde{E}_d{\rm D}^b\sigma^{cd}+ 2\varepsilon_{abc}\sigma^b{}_d{\rm D}^{\langle c}\tilde{E}^{d\rangle}- \tilde{E}^b{\rm curl}\sigma_{ab}\,,  \label{soG-B2}
\end{equation}
respectively. This set monitors the effects of gravitational radiation on electromagnetic signals propagating on a Minkowski background at second order.

According to expressions (\ref{soG-E2}) and (\ref{soG-B2}), the interaction between electromagnetic and gravitational radiation leads to electromagnetic waves that are generally different from their original counterparts. In particular, the specifics of the induced electromagnetic signals (i.e.~their amplitude, wavelength and direction of propagation) depend on those of the interacting sources. For our purposes, however, the most important implication of (\ref{soG-E2}), (\ref{soG-B2}) is that gravitational-wave distortions can drive electromagnetic signals. This is important, because forced oscillations provide the natural physical environment for resonances to occur. In other words, Eqs.~(\ref{soG-E2}) and (\ref{soG-B2}) imply that the interaction between the Weyl and the Maxwell field can lead to the resonant amplification of the latter.

\subsection{Resonant couplings}\label{ssRCs}
Resonances require rather special conditions, which makes them more likely to occur in controlled environments than spontaneously in nature. If gravito-electromagnetic resonances are possible, the system of (\ref{soG-E2}), (\ref{soG-B2}) should contain resonant solutions. To demonstrate (analytically) that this is indeed the case, let us ignore the backreaction of the electric component upon its magnetic counterpart and vice versa. Then, Eqs.~(\ref{soG-E2}) and (\ref{soG-B2}) reduce to
\begin{equation}
\ddot{E}_a- {\rm D}^2E_a= \sigma_{ab}\dot{\tilde{E}}^b+ 2\dot{\sigma}_{ab}\tilde{E}^b  \label{soG-E3}
\end{equation}
and
\begin{equation}
\ddot{B}_a- {\rm D}^2B_a= \sigma_{ab}\dot{\tilde{B}}^b+ 2\dot{\sigma}_{ab}\tilde{B}^b\,,  \label{soG-B3}
\end{equation}
respectively. To simplify the mathematics, without compromising the basic physics, suppose that both the electromagnetic signal and the gravitational wave are monochromatic. Then, we may use the decompositions
\begin{equation}
\tilde{E}_a= \tilde{E}_{(n)}\tilde{\mathcal{Q}}_a^{(1)(n)} \hspace{15mm} {\rm and} \hspace{15mm} \tilde{B}_a= \tilde{B}_{(n)}\tilde{\mathcal{Q}}_a^{(2)(n)}\,,  \label{foEM}
\end{equation}
for the components of the original electromagnetic field and
\begin{equation}
\sigma_{ab}= \sigma_{(k)}\mathcal{Q}_{ab}^{(k)}\,,  \label{foGW}
\end{equation}
for the gravitationally induced shear perturbation. As mentioned before, the characteristics of the gravitationally driven (the induced) electromagnetic wave depend on those of the originally interacting sources. This dependance is encoded in the decomposition
\begin{equation}
E_a= E_{(\ell)}\mathcal{Q}_a^{(1)(\ell)} \hspace{15mm} {\rm and} \hspace{15mm} B_a= B_{(\ell)}\mathcal{Q}_a^{(2)(\ell)}\,,  \label{soEM} \end{equation}
where $\mathcal{Q}_a^{(1)(\ell)}= \mathcal{Q}_{ab}^{(k)}\tilde{\mathcal{Q}}_{(1)(n)}^b$ and $\mathcal{Q}_a^{(2)(\ell)}= \mathcal{Q}_{ab}^{(k)}\tilde{\mathcal{Q}}_{(2)(n)}^b$ by construction. These are vector harmonic functions, with $\dot{\mathcal{Q}}_a^{(1)(\ell)}=0= \dot{\mathcal{Q}}_a^{(2)(\ell)}$, ${\rm D}^a\mathcal{Q}_a^{(1)(\ell)}=0={\rm D}^a\mathcal{Q}_a^{(2)(\ell)}$. Also,
\begin{equation}
{\rm D}^2\mathcal{Q}_a^{(1)(\ell)}= -\ell^2\mathcal{Q}_a^{(1)(\ell)} \hspace{15mm} {\rm and} \hspace{15mm} {\rm D}^2\mathcal{Q}_a^{(2)(\ell)}= -\ell^2\mathcal{Q}_a^{(2)(\ell)}\,,  \label{Qell}
\end{equation}
with
\begin{equation}
\ell^2= n^2+ k^2+ 2nk\cos\phi\,.  \label{ell}
\end{equation}
The last expression relates the wavelength of the gravitationally driven electromagnetic wave to the wavelengths and the interaction angle ($\phi$, with $0\leq\phi\leq\pi$) of the original sources. More specifically, given the values of $n$, $k$ and $\phi$, one can employ (\ref{ell}) to obtain a unique value for $\ell$.

Substituting decompositions (\ref{foEM}) -- (\ref{soEM}) back into expression (\ref{soG-E3}) and using the Laplace-Beltrami equation (\ref{Qell}a), we arrive at
\begin{equation}
\ddot{E}_{(\ell)}+ \ell^2E_{(\ell)}= \sigma_{(k)}\dot{\tilde{E}}^{(n)}+ 2\dot{\sigma}_{(k)}\tilde{E}^{(n)}\,.  \label{dsoG-E1}
\end{equation}
The driving force in the right-hand side is determined by the interaction between the original electromagnetic signal and the incoming gravitational wave. Following the discussion given in \S~\ref{ssEMWs} and \S~\ref{ssGWs}, these are monitored by
\begin{equation}
\tilde{E}_{(n)}= \mathcal{A}\sin(nt+\vartheta_\mathcal{A}) \hspace{15mm} {\rm and} \hspace{15mm} \sigma_{(k)}= \mathcal{B}\sin(kt+\vartheta_\mathcal{B})\,,  \label{lGWkEn}
\end{equation}
with the amplitudes and the phases depending on the initial conditions (see solutions (\ref{lEn}) and (\ref{lGWk})). Without loss of generality, we may set the phases in (\ref{lGWkEn}a) and (\ref{lGWkEn}b) to zero. Then, expression (\ref{dsoG-E1}) leads to
\begin{equation}
\ddot{E}_{(\ell)}+ \ell^2E_{(\ell)}= {1\over2}\,\mathcal{C}(n-2k)\sin\left[\left(k-n\right)t\right]+
{1\over2}\,\mathcal{C}(n+2k)\sin\left[\left(k+n\right)t\right]\,,  \label{dsoG-E2}
\end{equation}
where $\mathcal{C}=\mathcal{A}\mathcal{B}$. Clearly, proceeding in an exactly analogous way, we obtain the wave equation of the gravitationally driven magnetic component
\begin{equation}
\ddot{B}_{(\ell)}+ \ell^2B_{(\ell)}= {1\over2}\,\mathcal{C}(n-2k)\sin\left[\left(k-n\right)t\right]+
{1\over2}\,\mathcal{C}(n+2k)\sin\left[\left(k+n\right)t\right]\,.  \label{dsoG-B2}
\end{equation}
The last two equations show that the force driving the propagation of the gravitationally induced electromagnetic signal can be expressed as the superposition of two waves, the specifics of which are determined by the initial conditions. In particular, the effective wavenumbers of the driving oscillations are $m_{1,2}=k\mp n$, while that of the gravitationally driven electromagnetic signal ($\ell$) is given by (\ref{ell}). Clearly, when the initial conditions are favourable, these wavenumbers coincide. Then, both the driving and the driven waves will oscillate in resonance.

\subsection{Resonant solutions}\label{ssRSs}
The simplest resonant case occurs for $k=n$, that is when the original sources share the same wavelength (though their directions of propagation may differ). Then, it is straightforward to show that the pair of (\ref{dsoG-E2}), (\ref{dsoG-B2}) reduces to
\begin{equation}
\ddot{E}_{(\ell)}+ \ell^2E_{(\ell)}= \mathcal{F}\sin(mt) \hspace{15mm} {\rm and} \hspace{15mm} \ddot{B}_{(\ell)}+ \ell^2B_{(\ell)}= \mathcal{F}\sin(mt)\,,  \label{dsoG-EM1}
\end{equation}
with $\mathcal{F}=3\mathcal{C}k/2$ and $m=2k$. In other words, the driving force can be expressed as a single wave with amplitude and wavelength depending on those of the original sources. The solutions of (\ref{dsoG-EM1}a) and (\ref{dsoG-EM1}b) are respectively given by
\begin{equation}
E_{(\ell)}= \mathcal{W}\sin\left(\ell t+\varphi\right)+ {\mathcal{F}\over\ell^2-m^2}\sin\left(mt\right)  \label{soEell1}
\end{equation}
and
\begin{equation}
B_{(\ell)}= \mathcal{W}\sin\left(\ell t+\varphi\right)+ {\mathcal{F}\over\ell^2-m^2}\sin\left(mt\right)\,,  \label{soBell1}
\end{equation}
where the values of the integration constants ($\mathcal{W}$ and $\varphi$) are fixed at the onset of the gravito-electromagnetic interaction. Both of the above show the amplitude of the gravitationally driven (the induced) electromagnetic wave to diverge as $\ell\rightarrow m$. According to the relation (\ref{ell}), this corresponds to $\phi\rightarrow0$. Consequently, the interaction between electromagnetic and gravitational radiation, which share the same wavelength and propagate in the same direction on a Minkowski background, will resonantly amplify the electromagnetic signal.

Generally, the wavelengths of the interacting gravitational and electromagnetic signals will be different (i.e.~$k\neq n$). More specifically, in most physically realistic situations we expect that the wavelength of the gravitational radiation will far exceed that of its electromagnetic counterpart. This implies that typically $n\gg k$. In addition, the two originally interacting waves will generally have nonzero (and in principle different) phases. When $k\neq n$ and $\vartheta_{\mathcal{A}}$, $\vartheta_{\mathcal{B}}\neq0$, a similar analysis shows that the solution of Eq.~(\ref{dsoG-E1}) takes the form
\begin{equation}
E_{(\ell)}= \mathcal{W}\sin\left(\ell t+\varphi\right)+ {\mathcal{F}_1\over\ell^2-m_1^2}\sin\left(m_1t+\omega_1\right)+ {\mathcal{F}_2\over\ell^2-m_2^2}\sin\left(m_2t+\omega_2\right)\,, \label{soEell2}
\end{equation}
where $\mathcal{F}_{1,2}=\mathcal{C}(n\mp2k)/2$, $m_{1,2}=k\mp n$ and $\omega_{1,2}=\vartheta_{\mathcal{A}} \mp\vartheta_{\mathcal{B}}$. It goes without saying that an exactly analogous solution describes the propagation of the gravitationally driven magnetic component as well. Following (\ref{soEell2}), the amplitude of the electromagnetic wave diverges as $\ell\rightarrow|m_{1,2}|$. In the first case this translates to $\ell\rightarrow|k-n|$, which occurs when $\phi\rightarrow\pi$ (see Eq.~(\ref{ell})). When $\ell\rightarrow k+n$, on the other hand, we find that $\phi\rightarrow0$. Therefore, the interaction between electromagnetic and gravitational waves, which have different wavelengths but propagate in the same or in the opposite direction on a Minkowski background, will resonantly amplify the electromagnetic signal.

\section{Gravito-electromagnetic interaction in
cosmology}\label{sGE-MIC}
The basic physics leading to the Weyl-Maxwell resonances of the previous section are generic and not particular to the geometrical structure of the Minkowski space. This means that, at least in principle, analogous effects should also take place within more general spacetimes, like those related to cosmology for example.

\subsection{Background and linear dynamics}\label{ssBLDs}
In cosmology, the spacetime is no longer empty, static and flat, which means that the specifics of the interaction between the Weyl and the Maxwell fields generally change. Here, we will consider the gravito-electromagnetic coupling on an FRW background with Euclidean spatial hypersurfaces. We will assume, in particular, a spatially flat Friedmann universe containing a single barotropic fluid. The time evolution of this model is determined by the zero-order set
\begin{equation}
H^2= {1\over3}\,\rho\,, \hspace{10mm} \dot{H}= -H^2- {1\over6}\,\rho(1+3w) \hspace{10mm} {\rm and} \hspace{10mm} \dot{\rho}= -3H\rho(1+w)\,,  \label{bFRW}
\end{equation}
where $H=\dot{a}/a$ is the Hubble parameter ($a$ is the cosmological scale factor), $\rho$ is the density and $w=p/\rho$ is the barotropic index of the matter (with $p$ representing the isotropic pressure).

Let us now perturb the above defined background with propagating gravitational and electromagnetic waves.\footnote{The gravito-magnetic interaction was considered in~\cite{T1}, assuming a weakly magnetised FRW background. This meant that, strictly speaking, the mechanism was linear in perturbative terms. Here, to facilitate a direct comparison with the Minkowski-space study of the previous section, we adopt a magnetic-free background and consider the coupling between the Weyl and the Maxwell fields at the second perturbative level.} As before, the former are monitored by the transverse component of the shear, which now obeys the linear wave equation (e.g.~see~\cite{TCM})
\begin{equation}
\ddot{\sigma}_{ab}- {\rm D}^2\sigma_{ab}= -5H\dot{\sigma}_{ab}- {3\over2}\,H^2(1-3w)\sigma_{ab}\,.  \label{lddotsigma}
\end{equation}
At the same perturbative level, and given that our FRW background has flat spatial sections, the magnetic field decays adiabatically at all times. This means that, to linear order,
\begin{equation}
\tilde{B}_a\propto a^{-2}\,,  \label{lBa}
\end{equation}
irrespective of the type of matter that fills the universe and of its electric properties (i.e.~of whether the electrical conductivity is high or low~\cite{GR}). Next, we will look into the interaction between these two sources at the second perturbative level.

\subsection{Second order gravito-magnetic
interaction}\label{ssSOG-MI}
Inflation is probably the best available mechanism capable of causally producing large-scale gravitational and electromagnetic perturbations. These distortions start as subhorizon quantum fluctuations and freeze-out as classical perturbations once outside the Hubble scale (e.g.~see~\cite{KT}). Also, throughout the de Sitter phase, the cosmic medium is believed to be a very poor electrical conductor. After inflation, during reheating and the subsequent radiation era, the electrical conductivity of the matter increases rapidly. As a result, the electric fields gradually vanish and the currents freeze the magnetic field in with the fluid. Put another way, the post-inflationary universe satisfies the ideal-magnetohydronamic (MHD) requirements. Nevertheless, causality ensures that the MHD limit holds only well within the Hubble scale. Beyond the horizon, however, the electric currents had no time to establish themselves and the low-conductivity approximation of the de Sitter period still holds.

In what follows, we will consider the gravito-magnetic interaction at the time the Weyl and the Maxwell fields start crossing back inside the Hubble radius. Close to the horizon, the electric currents are not yet sufficiently strong to freeze the magnetic fields in with the matter. Thus, the electric fields are still present, though weakened by the ever increasing strength of the currents. All these mean that, near the Hubble threshold, the effect of the gravitational waves on the magnetic fields is monitored by the second-order expression~\cite{T2}
\begin{equation}
\ddot{B}_a- {\rm D}^2B_a+ 5H\dot{B}_a+ 3(1-w)H^2B_a= 2\left(\dot{\sigma}_{ab}+2H\sigma_{ab}\right)\tilde{B}^b\,.  \label{soddotB}
\end{equation}
In analogy with the Minkowski-space case discussed earlier, $B_a$ represents the gravitationally induced magnetic field. The main difference between the above and Eq.~(\ref{soG-B3}) in \S~\ref{ssRCs} is in the expansion and the matter terms seen in the left-hand side of (\ref{soddotB}). As in the Minkowski case, the most important feature (for our purposes) of the latter expression is the gravito-magnetic term in its right-hand side. The presence of this particular source-term means that Weyl-curvature distortions can drive magnetic oscillations in cosmological environments as well. In other words, Weyl-Maxwell resonances can also occur between cosmological gravitational waves and large-scale, primordial magnetic fields.

\subsection{Cosmological gravito-magnetic resonances}\label{CG-MRs}
The outcome of the gravito-magnetic interaction is determined by the coupled system of (\ref{lddotsigma})-(\ref{soddotB}) and by the evolution of our background model (see Eqs.~(\ref{bFRW})). To proceed further, we harmonically decompose the perturbations in a manner analogous to that used in \S~\ref{ssRCs}. Then, expressions (\ref{lddotsigma}) and (\ref{soddotB}) take the form
\begin{equation}
\ddot{\sigma}_{(k)}+ 5H\dot{\sigma}_{(k)}+ \left[{3\over2}\,(1-3w)H^2+\left({k\over a}\right)^2\right] \sigma_{(k)}= 0  \label{dlddotsigma}
\end{equation}
and
\begin{equation}
\ddot{B}_{(\ell)}+ 5H\dot{B}_{(\ell)}+ \left[3(1-w)H^2+\left({\ell\over a}\right)^2\right]B_{(\ell)}= 2\left(\dot{\sigma}_{(k)}+2H\sigma_{(k)}\right)\tilde{B}_{(n)}\,,  \label{dsoddotB}
\end{equation}
respectively. Note that $\ell$ is the eigenvalue/wavenumber of the gravitationally driven $B$-field, $k$ is that of the gravitational wave, $n$ determines the scale of the background $B$-field and the three are related by Eq.~(\ref{ell})~\cite{T1}. Also, according to the linear relation (\ref{lBa}), we have $\tilde{B}_{(n)}= \tilde{B}^0_{(n)}(a_0/a)^2$ at all times, where the zero suffix specifies a moment during the universe's evolution.

Current observations indicate that the current Hubble length lies close to $10^3$~Mpc. This implies that most of the astrophysically relevant scales have entered the horizon during the late radiation era. Then, solving (\ref{dlddotsigma}) for $w=1/3$, $a\propto t^{1/2}$ and $H=1/2t$, we obtain
\begin{equation}
\sigma_{(k)}= {\sqrt{k^2(\sigma_0^{(k)})^2 +a_0(\dot{\sigma}_0^{(k)})^2}\over k} \left({t_0\over t}\right)\cos\left[{k\over a_0H_0} \left(1-\sqrt{t\over t_0}\right)+ \theta\right]\,,  \label{rlsigma}
\end{equation}
with $\theta=\tan^{-1}(a_0\dot{\sigma}_0^{(k)}/k\sigma_0^{(k)})$ representing the phase and the zero suffix indicating the beginning of the radiation epoch. Substituting the above into the right-hand side of Eq.~(\ref{dsoddotB}), while taking into account that $\tilde{B}_{(n)}\propto a^{-2}$ always, we arrive at
\begin{eqnarray}
\ddot{B}_{(\ell)}+ {5\over2t}\,\dot{B}_{(\ell)}+ \left[{1\over2t^2}+ \left({\ell\over a_0}\right)^2\left({t_0\over t}\right)\right]B_{(\ell)}&=& {2\tilde{B}_0^{(n)}\sqrt{k^2(\sigma_0^{(k)})^2 +a_0^2(\dot{\sigma}_0^{(k)})^2}\over a_0}\, \left({t_0\over t}\right)^{5/2}\nonumber\\ &&\times\sin\left[{k\over a_0H_0} \left(1-\sqrt{t\over t_0}\right)+ \theta\right]\,.  \label{rdsoddotsigma}
\end{eqnarray}
The latter monitors the gravitationally induce magnetic field as it crosses inside the Hubble horizon of a radiation dominated, spatially flat FRW universe. Like its Minkowski counterpart before, expression (\ref{rdsoddotsigma}) describes a forced oscillation, where the driving force on the right-hand side is the result of the gravito-magnetic interaction. Unlike its Miknowski analogue, on the other hand, Eq.~(\ref{rdsoddotsigma}) contains extra terms due to the expansion of the universe and the presence of matter. Another difference is in the amplitude of the driving wave, which is no longer constant but decays in time.

The nature of (\ref{rdsoddotsigma}) means that resonant solutions, analogous to those found in Minkowski-space, are in principle possible in cosmology as well. The additional (expansion-related) terms, on the other hand, suggest that the specifics of the resonances may differ. Indeed, solving Eq.~(\ref{rdsoddotsigma}) we obtain (see also \S~IV.A and Eq.~(22) in~\cite{T1})
\begin{eqnarray}
B_{(\ell)}&=& \pm{\tilde{B}_0^{(n)}\sqrt{k^2(\sigma_0^{(k)})^2 +a_0^2(\dot{\sigma}_0^{(k)})^2}\over\ell H_0}\left({t_0\over t}\right){\rm Si}\left({\ell\mp k\over a_0H_0}\,\sqrt{t\over t_0}\right) \nonumber\\ &&\times\sin\left[{\ell\over a_0H_0}\,\left(\sqrt{t\over t_0}\mp{k\over\ell}\right)\mp\theta\right] \nonumber\\ &&\pm{\tilde{B}_0^{(n)}\sqrt{k^2(\sigma_0^{(k)})^2 +a_0^2(\dot{\sigma}_0^{(k)})^2}\over\ell H_0}\left({t_0\over t}\right) {\rm Ci}\left({\ell\mp k\over a_0H_0}\,\sqrt{t\over t_0}\right) \nonumber\\ &&\times\cos\left[{\ell\over a_0H_0}\,\left(\sqrt{t\over t_0}\mp{k\over\ell}\right)\mp\theta\right]\,,  \label{rsoB}
\end{eqnarray}
where ${\rm Si}(x)$ and ${\rm Ci}(x)$ represent the sine and the cosine integral functions respectively~\cite{AS}. Following the above, the gravitationally induced $B$-field generally oscillates with an amplitude that decays adiabatically in time (i.e.~$B_{(\ell)}\propto t^{-1}\propto a^{-2}$). At any given time, however, the amplitude depends on the initial conditions. This dependence is crucial because it also determines the argument of the aforementioned integral functions. The latter are known to satisfy the conditions $\lim_{x\rightarrow0}{\rm Si}(x)=0$ and $\lim_{x\rightarrow0}{\rm Ci}(x)=-\infty$~\cite{AS}. Therefore, the presence of the cosine integral function in solution (\ref{rsoB}), ensures that the amplitude of the gravitationally driven magnetic field diverges when the two sources oscillate in resonance (i.e.~for $k\rightarrow\ell$). The same integral functions also mean that resonances between the Weyl and the Maxwell fields in cosmology have a much narrower peak than their Minkowski space analogues. This is due to the universal expansion, which acts as a dumping force and weakens the overall effect of the gravito-electromagnetic resonance. Broader resonance peaks can be achieved, when the source term in the right-hand side of Eq.~(\ref{rdsoddotsigma}) drops slower with time. In particular, the resonant amplification of the induced $B$-field is exactly analogous to that seen in Eq.~(\ref{soBell1}), when the amplitude of the driving wave depletes as $t^{-2}$ instead of $t^{-5/2}$. This can happen, for example, when the background magnetic field decays as $a^{-1}$ (i.e.~$\tilde{B}\propto t^{-1/2}$ during the radiation era), like many of the $B$-fields proposed in various scenarios of primordial magnetogenesis~\cite{GR}. Then, one obtains
\begin{equation}
B_{(\ell)}= {2a_0\tilde{B}_0^{(n)}\sqrt{k^2(\sigma_0^{(k)})^2 +a_0(\dot{\sigma}_0^{(k)})^2} \over{k^2-\ell^2}}\left({t_0\over t}\right)\, \sin\left[{k\over a_0H_0}\left(\sqrt{t\over t_0}-1\right)-\theta\right]\,,  \label{rsoB1}
\end{equation}
for the gravitationally driven magnetic field~\cite{T1}. The above solution shows resonant magnetic amplification, when $k\rightarrow\ell$, analogous to that of the Minkowksi background (compare (\ref{rsoB1}) to Eq.~(\ref{soBell1}) earlier).

Note that, in principle at least, the amplification effect of the resonances seen in solutions (\ref{rsoB}) and (\ref{rsoB1}) is independent of the amount of energy stored in gravity-wave perturbations. In practise, this means that even the weak gravitational waves that inflation is expected to produced, can resonantly amplify primordial (most likely also inflationary) magnetic fields during the subsequent evolution of the universe. We should also point out that the amplification mechanism discussed above operates around the time of the second horizon crossing. There the accompanying primordial  electric fields are still present, though weakened by the ever growing effect of the currents. Deep inside the Hubble radius, however, the currents will dominate, eliminate the electric fields and freeze their magnetic counterparts into the highly conductive cosmic fluid. Once the freezing-in  process is completed, the ideal MHD limit will be established. Then onwards, resonances are no longer possible and our amplification mechanism ceases to operate~\cite{T1}. In the mean time, however, the universe has been permeated by large-scale magnetic fields that are substantially stronger than the original ones and, perhaps, strong enough to seed the galactic dynamo~\cite{M}.

\section{Discussion}\label{sD}
The basic physical principles behind the gravito-electromagnetic interaction are simple and essentially independent of the environment. More specifically, gravitational-wave distortions interact and can act as driving forces of electromagnetic radiation, irrespective of the structure of the host spacetime. This means that gravito-electromagnetic resonances are in principle possible both in astrophysics and cosmology, as well as in the laboratory. Nevertheless, there are still major differences between these environments, which are expected to affect the phenomenology of the resulting resonances. Among the goals of the present study was to identify such analogies and differences. The approach we have adopted describes gravitational-wave perturbations via the transverse component of the shear, rather than by using the pure-tensor part of the metric perturbation. This facilitates a direct and physically more transparent coupling between the gravitational and the electromagnetic fields through Maxwell's equations. In line with the purposes of our study, the Weyl-Maxwell coupling was considered first on an empty Minkowski background and the results were then compared to those obtained within a perturbed FRW cosmology. In either case, we have allowed for the presence of gravitational and electromagnetic waves and analysed the effects of the former on the latter at second order. In both cases, we have found that the system of equations monitoring the gravito-electromagnetic interaction contained resonant solutions. To be precise, under favourable initial conditions -- when the driving and the driven waves oscillate in tune, the amplitude of the gravitationally induced electromagnetic/magnetic field was found to diverge. There were, however, differences as well. Applied to Minkowski space, the gravito-electromagnetic interaction led to a textbook-like wave equation that describes forced oscillations free of any dumping terms. In cosmology, however, the expansion of the universe acts as a source of dumping that weakens the resonance, in the same (more or less) way that inhibits the growth of density perturbations. This weakening is reflected in the resonance peak, which in cosmology is generally narrower.

Resonances can occur either spontaneously in nature, or under controlled conditions in the laboratory. In either case, weak vibrations lead to oscillations of disproportionately large amplitude. It is not surprising therefore that resonances between gravitational and electromagnetic waves have been discussed as a potential gravity-wave detection mechanism. This way, the ubiquitous presence of electromagnetic radiation could prove a very useful ally in our ongoing effort to detect the still elusive gravity-wave signals. Cosmology is also an area where resonances between the Weyl and the Maxwell fields could have an effect, by amplifying weak primordial magnetic fields to astrophysically relevant strengths. Then, one may be able to address the still open issue of cosmic magnetogenesis, within the realm of standard cosmology and conventional electrodynamics. In principle, analogous gravito-electromagnetic resonances should also occur on spacetimes with more complicated structures, though then one would probably have to use numerical (rather than analytical) methods to identify them. Overall, in view of the generic nature of the Weyl-Maxwell coupling within general relativity and given that forced oscillations provide the natural stage for resonances to occur, the question may be how often such resonances take place, rather than whether they happen at all. This remains to be seen. If the answer turns out to be ``positive'', however, the ubiquitous presence of magnetic and electromagnetic fields in our cosmos may has been partly achieved at the expense of gravitational radiation.

\end{document}